\documentclass[utf8]{FrontiersinHarvard}

\usepackage{url,lineno,microtype}
\usepackage[hidelinks]{hyperref}
\usepackage[onehalfspacing]{setspace}
\usepackage{amsmath,amssymb,mathtools}
\usepackage{array}
\usepackage[hidelinks]{hyperref}
\usepackage{graphicx,xcolor}
\usepackage{cite}
\usepackage[font=footnotesize,labelfont=bf]{caption}
\usepackage{amsthm}
\usepackage{tcolorbox}
\tcbuselibrary{skins,breakable}
\usetikzlibrary{shadings,shadows}
\usepackage{multirow}
\usepackage[normalem]{ulem}
\nolinenumbers

\def\keyFont{\fontsize{8}{11}\helveticabold }
\def\firstAuthorLast{Acharya {et~al.}} 
\def\Authors{Gagan Acharya\,$^{1,*}$, Sebastian F. Ruf\,$^{2}$ and Erfan Nozari\,$^{1,3,4}$}
\def\Address{$^{1}$Department of Electrical and Computer Engineering, University of California, Riverside, CA, USA \\
$^{2}$Sustainability and Data Sciences Lab, Department of Civil and Environmental Engineering, Northeastern University, Boston, MA, USA \\
$^{3}$Department of Mechanical Engineering, University of California, Riverside, CA, USA \\
$^{4}$Department of Bioengineering, University of California, Riverside, CA, USA}
\def\corrAuthor{Corresponding Author}

\def\corrEmail{gacha002@ucr.edu}

\newcommand{\new}[1]{{ #1}}
\newcommand{\rem}[1]{}

\newcommand\sbf{\mathbf{s}}
\newcommand\ubf{\mathbf{u}}
\newcommand\vbf{\mathbf{v}}
\newcommand\wbf{\mathbf{w}}
\newcommand\xbf{\mathbf{x}}
\newcommand\ybf{\mathbf{y}}

\newcommand\Abf{\mathbf{A}}
\newcommand\Bbf{\mathbf{B}}
\newcommand\Cbf{\mathbf{C}}
\newcommand\Dbf{\mathbf{D}}

\newcommand\Kbf{\mathbf{K}}

\renewcommand\eqref[1]{Eq.~(\ref{#1})}

\begin{document}
\onecolumn
\firstpage{1}

\title[Brain Modeling for Control]{Brain Modeling for Control: A Review}

\author[\firstAuthorLast ]{\Authors} 
\address{} 
\correspondance{} 

\extraAuth{}

\maketitle

\begin{abstract}

Neurostimulation technologies have seen a recent surge in interest from the neuroscience and controls communities alike due to their proven potential to treat conditions such as epilepsy, Parkinson's Disease, and depression. The provided stimulation can be of different types, such as electric, magnetic, and optogenetic, and is generally applied to a specific region of the brain in order to drive the local and/or global neural dynamics to a desired state of (in)activity. For most neurostimulation techniques, however, an underlying theoretical understanding of their efficacy is still lacking. From a control-theoretic perspective, it is important to understand how each stimulus modality interacts with the inherent complex network dynamics of the brain in order to assess the controllability of the system and develop neurophysiologically relevant computational models that can be used to design the stimulation profile systematically and in closed loop. In this paper, we review the computational modeling studies of (i) deep brain stimulation, (ii) transcranial magnetic stimulation, (iii) direct current stimulation, (iv) transcranial electrical stimulation, and (v) optogenetics as five of the most popular and commonly used neurostimulation technologies in research and clinical settings. For each technology, we split the reviewed studies into (a) theory-driven biophysical models capturing the low-level physics of the interactions between the stimulation source and neuronal tissue, (b) data-driven stimulus-response models which capture the end-to-end effects of stimulation on various biomarkers of interest, and (c) data-driven dynamical system models that extract the precise dynamics of the brain's response to neurostimulation from neural data. While our focus is particularly on the latter category due to their greater utility in control design, we review key works in the former two categories as the basis and context in which dynamical system models have been and will be developed. In all cases, we highlight the strength and weaknesses of the reviewed works and conclude the review with discussions on outstanding challenges and critical avenues for future work.

\tiny
 \keyFont{ \section{Keywords:}
     computational modeling,
     system identification,
     neurostimulation,
     network control,
     deep brain stimulation,
     transcranial magnetic stimulation,
     electrical stimulation,
     optogenetics} 
\end{abstract}

\section{Introduction}\label{sec:intro}

Neurostimulation technologies have shown promising, initial success over the last 20 years in treating neurological disorders such as drug-resistant epilepsy (DRE), Parkinson's disease (PD), and psychological conditions such as depression~\citep{lin2017neurostimulation,ben2012neurostimulation,marangell2007neurostimulation}. Stimulation of the brain is also being increasingly used as a means to map the functional properties of various regions of the brain~\citep{mandonnet2010direct} and is also seen as a mode to enhance sensory-motor activity~\citep{jones2015longitudinal,toth2021effect}. While the potency of external electromagnetic stimulation has been established in the literature and neuromodulation under varying input conditions has been studied, clinical delivery of stimulation has largely been applied in an open-loop expert-driven manner where constant stimulation is provided for large periods of time (sometimes months). Such open-loop stimulation has been associated with inconsistent responses and sub-optimal modulation which can be linked to the high sensitivity of the brain to stimulation parameters~\citep{deeb2016proceedings}. Although closed-loop control of stimulation can potentially help address these issues (e.g., the RNS\textsuperscript\textregistered system from NeuroPace, Inc~\citep{skarpaas2019brain}), a large majority of works have focused on ON-OFF control that relies on medically-derived biomarkers such as signal line length, signal power in certain frequency bands, or tremor onset as opposed to predictive evaluation of the brain's response to stimulation.  These factors, together with the low efficacy of state-of-the-art controllers (e.g., complete seizure abatement in the case of the RNS\textsuperscript\textregistered system was reported in only 20\% of individuals) and the demand for more energy-efficient systems~\citep{ramirez2018evolving} make stimulation tuning via fully closed loop control a necessity.

\begin{table}[ht]
    \centering
    \setlength\extrarowheight{-1.5pt}
    \caption{List of common abbreviations used in this review.
    \vspace*{5pt}}
    \begin{tabular}{rlrl}
    \hline
       \textbf{fMRI}&Functional Magnetic Resonance Imaging &\textbf{EEG} &Electroencephalography\\
       \textbf{DBS}&Deep Brain Stimulation&\textbf{LFP}&Local Field Potential\\
       \textbf{DES}&Direct Electrical Stimulation&\textbf{ECoG}&Electrocorticogram\\
       \textbf{TMS}&Transcranial Magnetic Stimulation&\textbf{EMG}&Electromyography\\
       \textbf{tES}&Transcranial Electrical Stimulation&\textbf{MEP}&Motor Evoked Potential\\
       \textbf{ARX}&Auto-Regressive with eXogenous input &\textbf{ANN}&Artificial Neural Network\\
       \textbf{DCM}&Dynamic Causal Modeling&\textbf{MEG}&Magnetoencephalography\\
       \textbf{LSSM}&Linear State-Space Model &\textbf{FIR}&Finite Impulse Response\\
       \textbf{CBM}&Conductance-Based Model &\textbf{FEM}&Finite-Element Modeling\\
       \textbf{iEEG}&Intercranial EEG&\textbf{CV}&Cross Validation\\
       \textbf{BOLD}&Blood-Oxygen-Level-Dependent &\textbf{AIC}&Akaike Information Criterion\\
       \textbf{RMSE}&Root Mean Squared Error &\textbf{RSS}&Residual Sum of Squares\\\hline
    \end{tabular}
    \label{tab:abbrevs}
\end{table}

A major hurdle in the development of fully-closed-loop controllers is the mechanistic complexity of the brain and modeling its response to various forms of stimulation. Despite the rising relevance of neurostimulation and the rapid advancement in brain activity monitoring systems (e.g., fMRI, iEEG), the exact mechanism through which neurostimulation inputs interact with the brain connectome is still poorly understood. In this paper, we present a focused review of methods employed in the literature towards understanding the response of the brain to five of the most commonly used neurostimulation techniques, namely, deep brain stimulation (DBS), transcranial magnetic stimulation (TMS), direct electric stimulation (DES), transcranial electric stimulation (tES), and optogenetic stimulation. 

Existing reviews on different neurostimulation methods have largely focused on the efficacy of each method towards treating conditions such as PD, depression, or epilepsy and have had little emphasis on the brain modeling approaches that have been employed for each method~\citep{de2015neurostimulation,kassiri2017closed,magis2012advances,starnes2019review,schoenen2016noninvasive}. A few works have provided focused reviews of specific stimulation mechanisms~\citep{montagni2019optogenetics,chervyakov2015possible} based on the underlying neuroscience and usage of different biomarkers towards adaptive stimulation~\citep{bouthour2019biomarkers}. In \citet{lozano2019deep} and \citet{herrington2016mechanisms}, the authors have compiled various neural mechanisms (the inhibition of thalamic neurons, e.g.) explaining the DBS response to a range of disorders. To the best of our knowledge, existing reviews lack compiled collections of works on the modeling of the brain's response to various neurostimulation techniques, hence motivating the present review.
\begin{figure}[t]
    \centering
    \includegraphics[width=0.8\columnwidth]{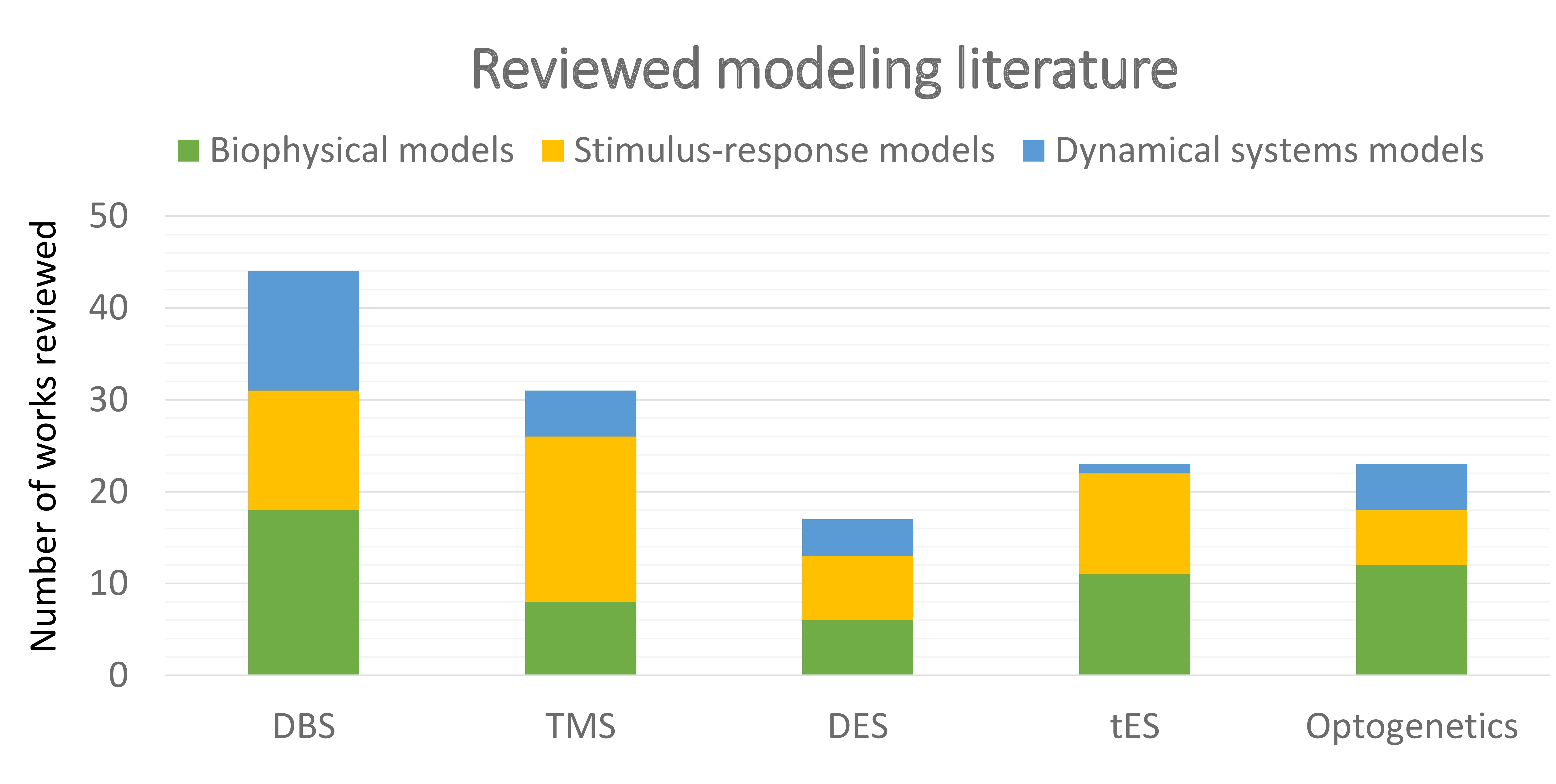}
    \caption{Distribution of literature reviewed in this paper}
    \label{fig:dist}
\end{figure}
Broadly, the computational modeling studies of the aforementioned neurostimulation methods can be categorized into three types of approaches\new{:}
\begin{enumerate}
    \item Works that use electric field equations and/or neuron models to build biophysically-derived models of the interaction between the stimulation input and the brain. These models are often parameterized by factors such as conductivity and geometry of brain tissues and are tuned to mimic observed data. As such, they are then commonly simulated to computationally optimize stimulation parameters such as input location and intensity.
    \item Works that use statistical and machine learning tools such as correlation, hypothesis testing, and/or artificial neural networks to model the overall stimulus-response profile of stimulation. Unlike the theory-driven nature of the first category, these models are fundamentally data-driven. However, they often are \emph{not} intended to capture the temporal dynamics of the brain's response to neurostimulation, as done by the last category.  
    \item Lastly, we have works where the impact of neurostimulation on the brain's network neural dynamics is learned using observed input-output time series data. In general, these methods do not make any assumptions regarding the underlying biophysics and rely mainly on data-driven algorithms. For simplicity of exposition, we will refer to this latter category as ``Dynamical System Models" while acknowledging the presence of dynamical system components in several of the works in the biophysical category.
\end{enumerate}

\begin{figure}[t]
    \centering
    \includegraphics[width=0.5\columnwidth]{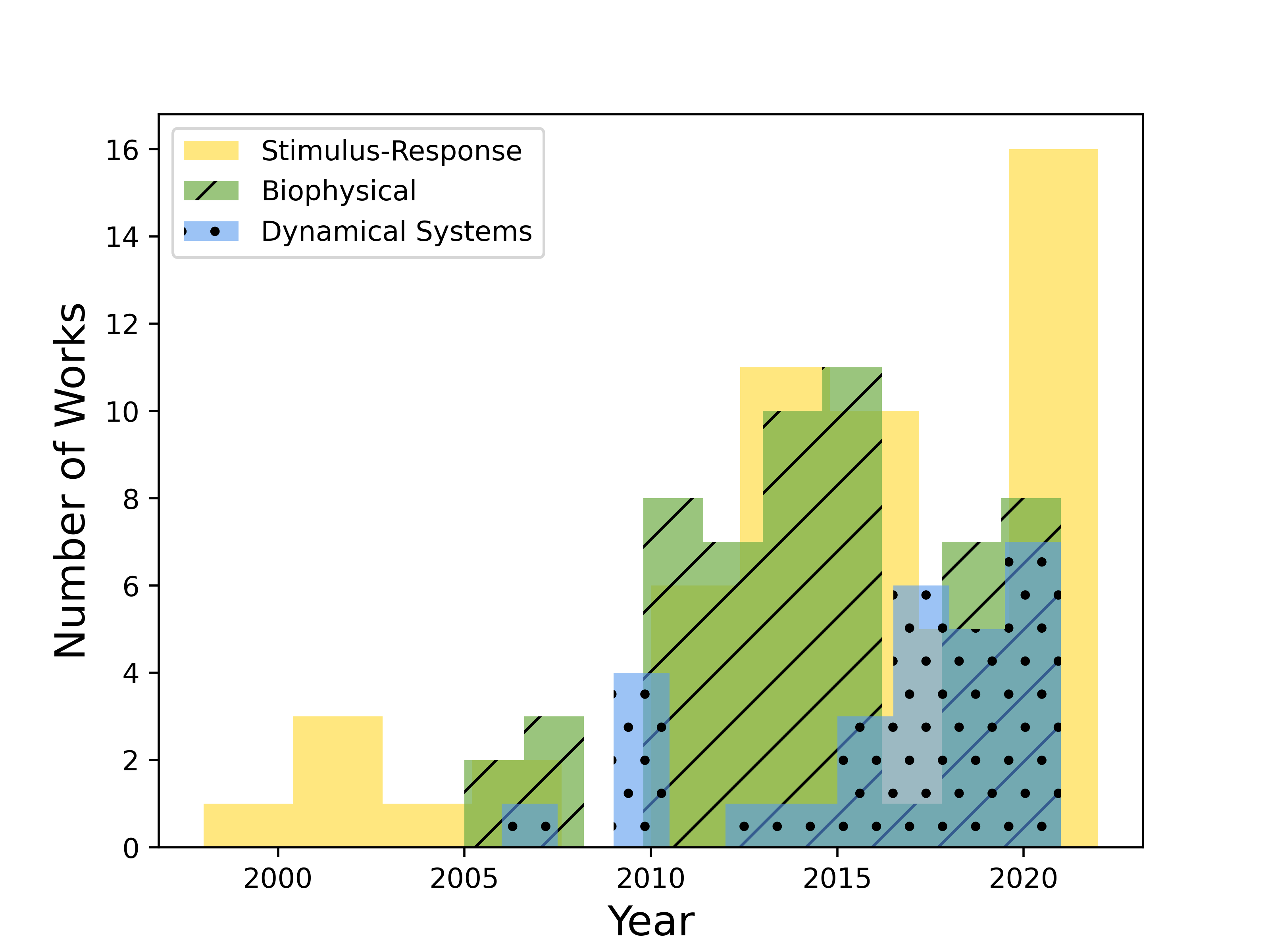}
    \caption{\new{\textbf{Year-wise spread of reviewed literature for the three types of approaches.} The use of data-driven dynamical systems modeling has markedly increased in recent years while biophysical and stimulus-response models have been employed for significantly longer times. Important to note is the fact that this figure is only intended to inform comparisons between categories, not the absolute historical frequency of the use of each category. The latter would require a historically uniform synthesis of the literature, particularly for biophysical and stimulus-response categories, as opposed to the focus of the present review on more recent works.}}
    \label{fig:temp_dist}
\end{figure}

\begin{figure}[t]
    \centering
    \includegraphics[width=0.8\columnwidth]{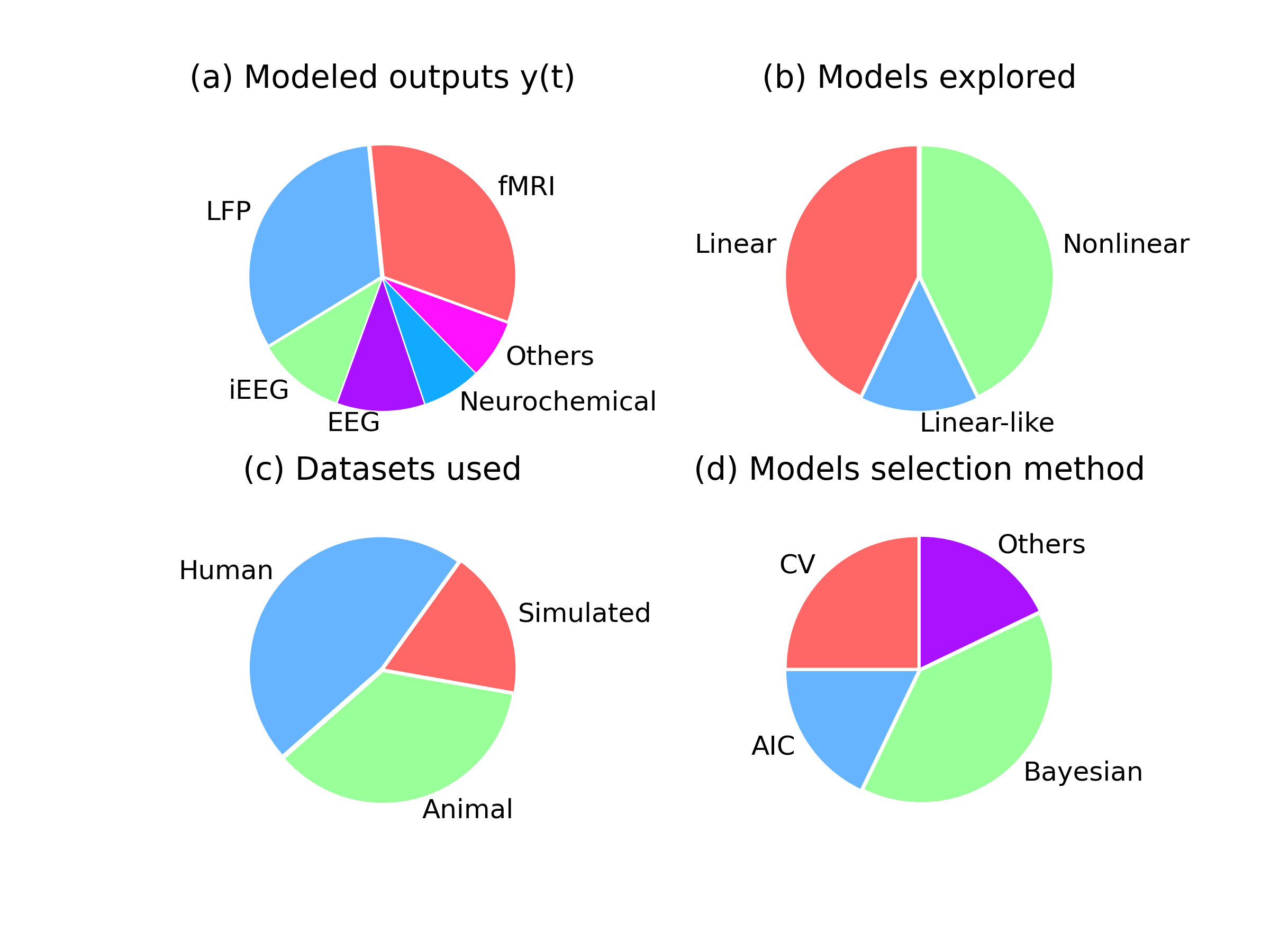}
    \vspace*{-20pt}
    \caption{\new{\textbf{Summary of dynamical systems models reviewed in this survey.} In panel (a), the distribution of modeled output $y(t)$ is plotted as a pie chart. DCM-based works have for the most part used fMRI output, while LFP/EEG/iEEG is more commonly used in works that use linear models like ARX/LSSM. In panel (b), we plot the proportion of model classes used in the modeling of brain dynamics under stimulation. Non-linear models include DCM, and ANN, while hybrid linear and Volterra models are grouped as linear-like. (c) Distribution of type/source of datasets used across reviewed works. In panel (d), we present various model order selection methods used in the literature.}}
    \label{fig:dyn_dist}
\end{figure}

In the following section, we provide an itemized review of the existing literature for each of the aforementioned five neurostimulation technologies and three methodological categories, see Figure~\ref{fig:dist} for a pictorial summary. While the first two categories have longer histories and more studies uniformly across all the reviewed stimulation technologies \new{as seen in Figures~\ref{fig:dist} and~\ref{fig:temp_dist}}, our particular focus in this review is on the dynamical systems models due to their applicability for therapeutic control design. \new{A summary of models in this latter category is illustrated in Figure~\ref{fig:dyn_dist} with breakdowns over output variables, model structures, datasets, and model selection criteria}. Nevertheless, we also review key studies and popular approaches in the biophysical and stimulus-response categories as they provide the \new{historical context} in which dynamical system models have been developed. \new{Finally, we review modeling works in the context of some of the most commonly studied brain disorders in order to facilitate interested readers' focused analyses of disease-specific literature.}

\new{\noindent\textbf{Review criteria}: The works presented in this review were largely obtained using keyword search from the Google Scholar database. Searched keywords were generally of the form ``\{stimulation\}+\{approach\}+\{specific model\}" where \{stimulation\} refers to one of the 5 neurostimulation techniques discussed in this paper, \{approach\} refers to one of three above-mentioned modeling approaches and \{specific model\} refers to commonly used modeling methods such as autoregressive models or neural networks. Our search provided us with a relatively larger number of literature for works concerning biophysical and stimulus-response models in comparison to dynamical system models. With the aim of keeping the review concise while investigating common modeling methods across literature, we have placed higher weight on more highly cited and recent works for inclusion in this review, while also carefully examining less cited and/or less recent works and including them if they provided sufficient methodological novelty or significance of findings compared to already-included references that would warrant an extended discussion. While the list of works reviewed here is not exhaustive, we hope to provide the reader with a broad and general understanding of prevailing methods that have been proposed for the modeling of the complex dynamics of the brain in response to neurostimulation. Finally, in the few cases where we refer to pre-prints which have not yet gone through peer-review, we indicate this with a $^\dagger$ mark next to the publication year of the respective articles.}

\section{Modeling the Brain's Response to Neurostimulation}\label{sec:Bu}

Consider a standard state-space model
\begin{align}\label{eq:ss}
\notag    \dot{\xbf}(t) \ \text{or} \ \xbf(t+1)&=f(\xbf(t), \ubf(t)) + \wbf(t),
    \\
    \ybf(t)&=h(\xbf(t), \ubf(t)) + \vbf(t),
\end{align}
where $\xbf(t)$, $\ubf(t)$, and $\ybf(t)$\footnote{\new{$\ybf(t)$ refers to the raw or processed recordings of the brain activity (e.g., EEG or fMRI ) observed during stimulation and its exact nature depends on factors such as stimulation mode (invasive vs non-invasive), region (e.g., sub-cortical or cortical) and process of interest (e.g., electrical vs chemical changes).}} denote the system's state, input, and output and $\wbf(t)$ and $\vbf(t)$ denote process and measurement noise.
\emph{Network control models} of this form have been used extensively, particularly over the past decade, to study various aspects of brain function and dynamics (see, e.g., ~\citep{ZH-SVS:17,nozari2020brain,gu2015controllability,kenett2018driving,schiff2011neural,becker2018large,yang2021modelling,singh2020estimation,pedoto2010system,nozari2020hierarchical2}).
Arguably, the most distinctive feature of these models compared to the classical \emph{dynamical systems models} of the brain is the presence of the control input $\ubf(t)$.%
\footnote{Another distinctive feature, though not relevant here, is the presence of an output variable $\ybf(t)$ that can be in general different from the \emph{latent} states $\xbf(t)$, see, e.g.,~\citep{ZH-SVS:17}.}
However, the precise physiological meaning of the control term $\ubf(t)$ and its ties both to the brain's intrinsic dynamics as well as to the measurable waveforms of neurostimulation are neither adequately understood nor a matter of consensus.
In this section, we focus on neurostimulation and review recent research and progress in characterizing the meaning of the control input $\ubf(t)$ and its impact on the local and global dynamics of the brain in the context of the most widely studied neurostimulation technologies listed in Section~\ref{sec:intro}.

\subsection{Deep Brain Stimulation}\label{sec:dbs}

Deep Brain Stimulation (DBS) is one of the oldest and most effective neurostimulation technologies with applications in various neuropsychiatric disorders~\citep{ramirez2018evolving}, particularly in treating movement disorders~\citep{wichmann2016deep} such as Parkinson's disease~\citep{ligaard2019deep}. DBS involves the (invasive) placement of electrodes that penetrate to varying depths inside the brain and injection of electrical currents to the surrounding brain tissue using one or more contacts on the implanted electrode~\citep{perlmutter2006deep}. In this section, we first briefly review some of the key studies on the modeling of the low-level biophysics as well as the stimulus-response characteristics of DBS and then provide a more in-depth account of the data-driven dynamical system modeling studies thereof.

\subsubsection{Biophysical Models}\label{sec:dbs-1}

The eye-catching success of DBS has led to significant computational research, both to understand its mechanisms of action and to optimize them~\citep{herrington2016mechanisms}. 
A common approach has been a combination of finite element modeling to estimate the electrical field distribution in the tissue surrounding the DBS electrode and 
conductance-based neuron modeling
simulated under the resulting electrical field to test whether the neurons in any given location will be sufficiently depolarized to generate an action potential~\citep{teplitzky2016model,keane2012improved,zitella2013computational,birdno2012stimulus,anderson2020neural,anderson2019anodic,van2018avoiding}. In a (single-compartment) conductance-based model (CBM), each neuron is modeled as a capacitive element~\citep{PD-FLA:01,CK:04}
\begin{align}\label{eq:cbm}
    C_i \frac{d}{d t} v_i(t) &= I_i(t) + \sum_{j = 1}^n \bar g_{ij} p_{ij}(t) [E_j - v_i(t)], \quad i = 1, \dots, n,
\end{align}
where $v_i$ is the membrane potential of neuron $i$, $C_i$ is its membrane capacitance, $I_i = I_i(v_i)$ denotes the total \emph{internal} voltage-dependent currents of this neuron (responsible for creating nonlinear spiking activity), $\bar g_{ij}$ is the maximal conductance of the synapse from neuron $j$ to neuron $i$ (with $\bar g_{ij} > 0$ for excitatory synapses, $\bar g_{ij} < 0$ for inhibitory ones, and $\bar g_{ij} = 0$ for disconnected neuron pairs), $p_{ij}(t)$ is the time-dependent probability that the channels in this synapse are open (most strongly dependent on the time of neuron $j$'s action potentials), and the constant $E_j$ is the reversal potential of the synapse.%
\footnote{Multi-compartment CBMs (more common in the biophysical modeling of DBS) are simple extensions of~\eqref{eq:cbm} where each neuron is instead modeled as a connection of multiple capacitive elements with different geometric orientations. Each compartment is then modeled as a capacitive element as in~\eqref{eq:cbm} but with additional terms representing the electrical flows between connected compartments.}
In the absence of DBS, each neuron will spike whenever its $v_i(t)$ increases beyond a threshold. Depending on the geometric morphology of each neuron, the electrical field created by the DBS will then increase (de-polarize) or decrease (hyper-polarize) this autonomous potential, thus increasing or decreasing its likelihood of spiking.

The complexity of this approach has motivated a number of simplifications, including the elimination of the CBM 
\citep{buhlmann2011modeling,martens2011spatial,janson2020activation,nguyen2019analysis}. 
These works still estimate the electrical field distribution, but use a simple and fixed threshold for neuron activation, either from prior experimental implementations of DBS or from a comparison with CBMs.


A further simplification, and a rather critical one, comes from works that assume a ``naive model" of how DBS affects the input current of nearby neurons~\citep{popovych2012desynchronizing,lysyansky2011desynchronizing,brocker2013improved,gorzelic2013model}. We refer to a dynamical model of neurostimulation as \emph{naive} when differential/difference equations describing autonomous neural dynamics are simply augmented by an extra additive term which directly equals (or is proportional to) the applied stimulation waveform. In the case of a CBM, e.g., \eqref{eq:cbm} changes to
\begin{align*}
    C_i \frac{d}{d t} v_i(t) &= I_i(t) + \sum_{j = 1}^n \bar g_{ij} p_{ij}(t) [E_j - v_i(t)] + u(t)
\end{align*}
where $u(t)$ is taken to be equal (or proportional) to the applied DBS waveform.
Clearly, these idealized models are quite generic and almost equally applicable to any stimulation technique. In return, however, these works are able to examine beyond the activation of the tissue near the DBS electrode and study the local network effects of DBS, such as desynchronizing reset modulation~\citep{popovych2012desynchronizing,lysyansky2011desynchronizing}, temporally non-regular stimulation~\citep{brocker2013improved}, and closed-loop feedback stimulation~\citep{gorzelic2013model}, as well as the optimization of the waveform shape of the DBS stimulation using a genetic algorithm~\citep{brocker2017optimized}.

\citet{grant2012simulation} take this one step further, combining electrical field calculations with CBMs of the subthalamic nucleus (the main stimulation region for the treatment of PD) and mean-field models of a handful of other subcortical and cortical regions. While a valuable extension, the considered network is still far from a complete connectome. The importance of the latter and global network effects have indeed been shown at least in the response of patients with treatment-resistant depression to DBS~\citep{riva2018connectomic} and is perhaps relevant for most psychiatric disorders. To account for this, \citet{riva2018connectomic} combine estimations of the activated tissue similar to the ones above together with region of interest (ROI) tractography to find the network of white matter connecting to the tissue activated by DBS and use that to plan surgical targeting of the DBS electrode.

Overall, the bottom-up approach of biophysical models inevitably sacrifices scalability in return for maximum biological interpretability. Indeed, large-scale interconnections of biophysical models are possible~\citep{amunts2016human}, but they require randomly distributed synaptic weights ($\bar g_{ij}$ in~\eqref{eq:cbm}) which in turn take away from the model's biological realism.
These and other issues with bottom-up modeling has motivated works that directly use the observed DBS data to study the underlying stimulation dynamics in a top-down manner, as reviewed in Sections~\ref{sec:dbs-2} and~\ref{sec:dbs-3}.


\subsubsection{Stimulus-Response Models}\label{sec:dbs-2}

A commonly used approach to study the modulation achieved by DBS is via the statistical analysis of stimulation effects on collected brain measurements like local field potentials (LFPs) or intracranial EEG (iEEG) \citep{marceglia2007basal, swann2011deep}. Unlike biophysical models, these works 
use statistical methods to locate regions and periods where DBS has had an effect on the measurement data. In works like \citep{rosin2011closed}, this is done by cross-correlating the DBS input waveform with LFPs to find channels where the effects are significant. However, the most widely used technique involves the computation of biomarkers~\citep{bouthour2019biomarkers} such as \new{beta-band power (signal power in the 13-30Hz range), high gamma oscillations (60–90Hz), and tremor bursts (4–6Hz)} from the measured data and the comparison of biomarker values corresponding to periods with no DBS to those with stimulation \citep{saenger2017uncovering, swann2011deep}. The exact biomarker that needs to be used has been a matter of debate for decades \citep{little2012brain, bouthour2019biomarkers} and it varies based on the disease and the DBS application.

One of the main research objectives concerning neurostimulation is the development of Brain-Computer Interfaces (BCIs), where the goal is to process brain measurements in real-time in order to obtain stimulation-relevant biomarkers and decide about stimulation parameters in an adaptive manner. The majority of the existing works regarding adaptive DBS (aDBS) are based on the comparison of computed biomarkers with a clinically obtained threshold \citep{hoang2017biomarkers, beudel2016adaptive, little2013adaptive,priori2013adaptive}. While these methods have been shown to perform better than open-loop continuous DBS, biomarkers are still only manually-selected low-dimensional representations of the brain activity that may not fully capture the DBS effects.  Further, the adaptation achieved in these works is limited to controlling DBS durations in an on/off manner and requires expert intervention to adjust other parameters like frequency and amplitude. Motivated by these, data-driven machine learning methods have been employed in \citep{shukla2012neural,khobragade2015towards} and have been shown to perform better than biomarker-based aDBS. In both these works, an artificial neural network (ANN) was used to predict tremors using surface-electromyogram (sEMG) data which was then used to decide the application of DBS input. However, these works are still oblivious to the rich temporal dynamics of DBS response, as reviewed next.

\subsubsection{Dynamical System Models}\label{sec:dbs-3}

More recently, a number of works have approached the problem of DBS modeling from a completely different, system identification approach (Table \ref{tab:dbs}). In~\citep{liu2016closed}, the authors carry out an input-output black-box system identification to fit an autoregressive with exogenous input (ARX) model of the form in \eqref{eq:arx} which takes the DBS signal as the input $u(t)$ and produces the basal ganglia-to-thalamus synaptic conductance as the output $y(t)$. 
The resulting model is then used to optimize the DBS input via the generalized predictive control framework. Similarly, \citep{santaniello2010closed} also fits an ARX model that takes the DBS signal as the input, but they instead choose the DBS-generated LFP as the model output.
\begin{align}\label{eq:arx}
    y(t+1) &= \sum_{j=0}^{p-1}a_jy(t-j)+\sum_{k=0}^{m-1}b_ku(t-k)+\wbf(t)
\end{align}The authors generate the LFP data by applying a range of stimulation frequencies/amplitudes and the resulting model is then used together with minimum variance control to shape the power spectral density of the LFP. Even though both works train their ARX models on simulated rather than experimental data, their pragmatic and data-driven methodology can be valuable in mitigating the complexity of the DBS mechanism. The latter is in turn achieved by focusing only on the end-to-end, input-output mapping from tunable stimulation parameters to neural biomarkers of interest in any disorders, or even behavioral measurements similar to those studied in~\citep{medvedev2019control}.

\begin{table}[ht]
    \centering
    \resizebox{\textwidth}{!}{%
    \begin{tabular}{|l|l|l|l|l|l|l|}
    \hline
       Model & Data & Paper & $y(t)$ & $u(t)$& \new{Performance} &Order \\
        &  &  &  & & \new{measure} &Selection \\
       \hline\hline
       \multirow{5}{*}{ARX}&\multirow{2}{*}{Simulated}&
        \citet{liu2016closed} & Conductance & Stim. current &RMSE &CV  \\ \cline{3-7}
        &&\citet{santaniello2010closed} & LFP & Stim. current &RMSE &CV \\ \cline{2-7}
        &\multirow{3}{*}{Animal}&\citet{wang2017proof} & LFP power &ON/OFF & $R^2$&CV  \\\cline{3-7}
        &&\citet{behrend2009toward} & Glutamate& ON/OFF & $R^2$& AIC \\ \cline{3-7}
        &&\citet{pedoto2010system} & LFP&  Stim. current & -& Literature  \\\hline
        ANN&Animal&\citet{trevathan2017computational} & Dopamine& Stim. current  &$R^2$ & CV \\ \hline
        Volterra&\multirow{2}{*}{Simulated}& \citet{su2018nonlinear}& LFP &Stim. current & RSS & CV\\\cline{3-7}
        Kernels&&\citet{chang2020model}& LFP &Stim. current& MSE &CV\\\hline
        Hybrid ARX&Human&\citet{haddock2017model}& LFP & Stim. amp. &- &- \\\hline
        \multirow{3}{*}{DCM}&\multirow{3}{*}{Human}&
        \citet{kahan2014resting} & fMRI & - & Posterior &Bayesian  \\ \cline{3-7}
        &&\citet{kahan2019deep} & fMRI & -  &Posterior &Bayesian \\  \cline{3-7}
        &&\citet{gibson2017impact} & fMRI & -  & Posterior&Bayesian \\  \cline{3-7}
        &&\citet{van2018generic} & LFP/MEG & - & -&Bayesian \\ \hline
    \end{tabular}%
    }
    \caption{Summary of data-driven dynamical system models for DBS. }
    \label{tab:dbs}
\end{table}
As far as learning parametric input-output models from real-world data is concerned, most of the existing DBS literature focuses on non-human subjects, including rodents \citep{wang2017proof,behrend2009toward}, rhesus macaques \citep{pedoto2010system}, or swines \citep{trevathan2017computational}. In \citep{pedoto2010system}, the authors extend the work in \citep{santaniello2010closed} to data collected from non-human primates (NHPs) via an experimental design that involved varying the stimulation frequency randomly while keeping the amplitude fixed. Although the model learned from this design could not generalize as well as the one in \citep{santaniello2010closed}, it is interesting to note that a relatively low-order scalar model was able to explain a large portion of the variance seen in the real data.

In \citep{wang2017proof}, the authors \new{train} an ARX model with the input being the application of DBS (binary on/off) and the output being an LFP-based psychological marker sampled on a daily basis. Unlike in \citep{santaniello2010closed} where the learned model was used to control the DBS current in real-time, the model obtained here was used to design a fuzzy logic-based controller to decide the application of DBS once every day based on the biomarker history. Application of ARX modeling has also been demonstrated in \citep{behrend2009toward} for the prediction of the extracellular glutamate (the brain's main excitatory neurotransmitter) level in rat brains as the output response. The success of linear models such as ARX in modeling what is inherently a nonlinear process is a subject of research and has been attributed to the brain's inherent spatio-temporal averaging processes, among other reasons \citep{nozari2020brain}.

Nonlinear models such as AR-Volterra Kernels \citep{su2018nonlinear} and ANNs \citep{trevathan2017computational} have also been used for the modeling of DBS response.
In \citep{chang2020model}, the authors take this one step further by learning a nonlinear ARMA-Volterra model and providing a predictive control algorithm based thereon. While these nonlinear models and the linear models discussed earlier are demonstrated to achieve good accuracy in predicting DBS response \new{(e.g., $R^2=0.99$ in \citep{behrend2009toward} and normalized MSE of $0.07$ in \citep{chang2020model})}, they are still scalar models where the output of any brain region/channel is modeled independently of the measurements at other regions/channels. Hence, these models have limited utility in  understanding how DBS affects the brain's large-scale network dynamics and the connectivity between different regions. 
Furthermore, these models' lack of network interactions has made it inevitable to give the model of all regions direct access to the DBS input history, as opposed to the more biologically realistic scenario where input enters into the dynamics only through the stimulation site. 

Combining the power of data-driven modeling with the interpretative nature of biophysical models is an ongoing area of research \citep{rowe2004estimation,dura2019netpyne,freestone2014estimation}. Perhaps dynamic causal modeling (DCM) originally proposed in \citep{friston2003dynamic} has been the most leveraged modeling framework towards this objective. While the exact formulation of DCM can vary widely depending on the stimulation type and the measurement modality, it is ultimately a special case of the model in \eqref{eq:ss} where $f(.)$ often has a bi-linear form for fMRI outputs and (sigmoidal) mean-field forms for EEG and MEG outputs and external inputs (neurostimulation in this case) ``modulate" (increase or decrease in an affine form) the interaction weights between the states (corresponding to the activity of modeled brain regions). In the case of fMRI outputs, for instance, the dynamics take the form 
\begin{equation}
    \begin{aligned}
    \dot \xbf(t) &= (\Abf + u(t) \Bbf) \xbf(t)
    \\
    \ybf &= h(\xbf)
\end{aligned}
\end{equation}

where $u(t)$ is the neurostimulation input and $h(\cdot)$ models the hemodynamic (for fMRI) or electromagnetic (for EEG and MEG) processes by which the neuronal states $\xbf(t)$ give rise to the recorded neuroimaging time series. Note that $h(\cdot)$ is often a dynamic process, hence the used notation.
We point interested readers to \citep{marreiros2010dynamic,stephan2010ten} for more details on the DCM theory and learning. While DCM models are learned from data and can potentially be used for feedback control, it is largely seen as a tool to examine the changes in effective connectivity (i.e., $(\Abf + u(t) \Bbf)$) between brain regions when comparing the stimulus-on and stimulus-off conditions. In the majority of works involving DCM and neurostimulation (e.g.,~\citet{kahan2014resting}), a set $\mathcal{M}=\{m_1,m_2,...,m_M\}$ of biophysically plausible network models are initially proposed (corresponding to different sparsity patterns in $\Abf$ and $\Bbf$), all fit to data using the standard Bayesian paradigm of DCM, and the configuration with maximum model evidence (i.e., $\arg\max_m p(\ybf|m)$) is selected.
This model $m$ is then used to identify the effect of stimulation on the connections between the selected regions of interest. Such analyses have been done for DBS with fMRI data in works such as \citet{kahan2014resting,kahan2019deep,gibson2017impact} and with LFP/MEG recordings in \citet{van2018generic}. Note that, as described, these works learn separate models for stimulation-on and stimulation-off durations and do not consider stimulation as a driving input to network nodes. 

A similar split-modeling approach has also been employed in the context of autoregressive models in \citep{haddock2017model}, where hybrid ARX modeling was used to model tremor power obtained from LFP recorded during the stimulation of human brain. The authors here trained an AR and ARX model by separating the data into DBS-off and DBS-on durations, respectively. The learned hybrid model was then used in a simulation to develop a predictive controller with the goal to optimize the DBS effects while minimizing power consumption. Despite its limitations, a valuable aspect of this work lies in its 
demonstration of model effectiveness for closed-loop DBS control which we believe is a critical factor to be considered in addition to cross-validated predictive accuracy when assessing the quality of learned dynamical models. 



\subsection{Transcranial Magnetic Stimulation}

Transcranial Magnetic Stimulation (TMS) is a non-invasive technology whereby neuronal tissue is stimulated by the generation of a focal magnetic field using one or more electromagnetic coil(s) placed near the scalp~\citep{hallett2007transcranial}. In turn, the time-varying magnetic field creates an electric field that can de/hyper-polarize (excite/inhibit) neurons.

Throughout the decades-long history of TMS, various stimulation waveforms with distinct effects have been devised, the most notable of which are single-pulse, paired-pulse, and repetitive TMS~\citep{galletta2011transcranial}. Unlike DBS which is often believed to activate all neurons in the volume of tissue activated (VTA), TMS has long been recognized to have a potential for differentially activating excitatory or inhibitory interneurons in its VTA~\citep{beynel2020effects}. Such differential effects have been recognized both for paired-pulse TMS, with short inter-pulse internals ($<5$ms) giving rise to local inhibition and long inter-pulse intervals ($>5$ms) resulting in local excitation~\citep{mishra2011repetitive}, as well as for repetitive TMS (rTMS), where low-frequency and high-frequency rTMS are widely believed to give rise to local inhibitory and excitatory effects, respectively~\citep{guse2010cognitive,beynel2020effects}. The threshold for switching from inhibitory to excitatory (ranging between $1-5$Hz) and the mechanisms responsible for such frequency dependence are nevertheless still a matter of debate~\citep{luber2016application,beynel2020effects}. Nevertheless, from a network controls perspective, the potential for having a differentially excitatory or inhibitory effect (both positive and negative $\ubf(t)$) provides a remarkable advantage for TMS over DBS and many other stimulation technologies.
Without such a direct means of inhibition, network control can still be possible but more sophisticated control algorithms would be required to indirectly create inhibition, e.g., using excitation of local inhibitory interneurons.

In the following and in parallel to Section~\ref{sec:dbs}, we review the literature on the biophysical, stimulus-response, and dynamical system modeling of TMS across these stimulation modalities.

\subsubsection{Biophysical Models}

A major body of work has sought to model the effects of TMS on neural activity via calculations of the electric field that results from TMS. Electric field calculations are often performed via finite-element modeling (FEM)~\citep{thielscher2015field,salinas2007detailed} and can be performed at varying levels of accuracy or simplification.
Once the electric field is obtained across space and time, a standard approach consists of simulating the response of CBMs such as those in~\eqref{eq:cbm} at various positions/orientations to the electric field~\citep{salvador2011determining,de2016modeling,goodwin2015subject,gomez2020review,shirinpour2021multi}. Similar to the biophysical analyses for DBS, this is often achieved by first transforming electric field to electric potential (by integrating $E = -\nabla V$ over space) and then adding the resulting potential to the neuron's \emph{extracellular potential}. This is often performed either directly using the cable equation~\citep{salvador2011determining,de2016modeling}
or using multi-compartmental CBMs which essentially provide a spatial discretization of the cable equation to simplify the partial differential equation (PDE) thereof into multiple ordinary differential equations (ODEs)~\citep{goodwin2015subject}.

In comparison to naive models defined in Section~\ref{sec:dbs-1}, the effect of TMS using these field-based calculations also boils down to a simple, additive term in a CBM. Unlike a naive model, however, the waveform of this additive term is not simply the waveform of the TMS pulse, but rather the result of the detailed FEM and electromagnetic transformations. Nevertheless, naive models have also been used for the computational modeling of TMS~\citep{alagapan2016modulation}, albeit less often than DBS due to the more indirect mechanisms of action of the TMS.

Also similar to, but distinct from, the naive models is the framework of~\citep{esser2005modeling} where the effect of TMS is modeled directly (without field calculations) by a change in the synaptic conductance of neurons within the TMS focus. 
A second-order synaptic dynamics of the form (cf.~\eqref{eq:cbm})
\begin{align}\label{eq:ltiir}
    p_{ij}(t) \propto -e^{t/\tau_1} + e^{t/\tau_2}
\end{align}
is also included, where $\tau_1, \tau_2$ are the synapse's rise and decay time constants, further departing from the naive models and increasing biophysical realism.

\subsubsection{Stimulus-Response Models}

Most of the existing closed loop controllers of TMS are based on feedback from surface measurements like EEG, EMG \citep{tervo2020automated,tervo2022closed,kraus2016brain} or imaging modes like functional magnetic resonance imaging (fMRI)~\citep{jung2020modulating,hernandez2013identifying}. 
Some of the earliest studies of neural response to TMS are based on the changes in the motor evoked potentials (MEPs) \citep{peterchev2013pulse,pascual1998study}. MEPs refer to EMG signals recorded from muscle activity in response to stimulation of the respective area in the motor cortex and constitute one of the major paradigms for studying the neural effects of TMS. The aforementioned works typically attempt to characterize the TMS response via input-output plots under different stimulation parameters such as intensity and duration which could then be used for functional mapping or TMS control.

TMS-induced blood-oxygenation-level-dependent (BOLD) changes as observed by fMRI is another popular method of analysis due to its spatial resolution (millimeter range) \citep{jung2020modulating}. However, its use for real-time control of TMS is limited due to its low sampling rate (seconds range) and the fact that imaging techniques capture blood flow changes and not direct neuronal activity induced by stimulation. More recently EEG has gained great attention as a tool to study TMS-evoked transient changes \citep{bonato2006transcranial, iramina2002effects,kahkonen2005prefrontal, komssi2002ipsi}. This is largely influenced by the high temporal resolution (millisecond range) associated with EEG and the consistency of the EEG responses to TMS \citep{bonato2006transcranial}. Nonetheless, these works still often provide qualitative templates of the brain's regional responses to TMS, leaving the generative mechanisms through which the input interacts with the connectome unexplored.

Finally, a rather large body of work has explored the power of deep neural networks for computing the electrical activity induced by TMS in the brain \citep{yokota2019real,afuwape2021neural,sathi2021deep,yarossi2019experimental,akbar2020mapping}. While these works can be differentiated based on the exact specification of input and output variables, most of these models can be better categorized based on the direct or indirect usage of magnetic resonance (MR) images along with the coil parameters. The indirect approach involves the processing of MR images to compute anatomic brain models, which are then used along with the TMS parameters to simulate the E-field distribution. This in turn is fed into to a deep neural network to estimate the MEP~\citep{akbar2020mapping, yarossi2019experimental} or activation thresholds~\citep{aberra2022rapid}. While using processed MR images can help simplify the modeling (by requiring smaller network size and lesser data, for example), this step is often computationally expensive and relies on conduction models that are manually tuned. In contrast, MR images and coil parameters have been used directly as inputs to a convolution neural network to determine either the whole-brain electric field distribution~\citep{yokota2019real} or the electric field statistics such as maximum induced electric field~\citep{afuwape2021neural}. While the models in the direct approach were trained and tested using simulated TMS data, the predictability achieved with these models combined with the computational performance that comes with not having to pre-process MR images still provides a case for end-to-end data-driven modeling.

\subsubsection{Dynamical System Models}

Recent works have employed a data-driven approach to dynamical system modeling for TMS~\citep{chang2019assessing,kiakojouri2020brain}, where linear state space models (LSSM) were used to fit EEG response to TMS. In essence, LSSMs of the form shown in \eqref{eq:lssm} are learned using subspace identification methods. However, \citet{chang2019assessing} instead parameterize their state equation using a \new{linear multivariable ARX (MVARX)} style model, which is then trained using an expectation-maximization style algorithm.
In this work, it was observed that fully-connected LSSMs perform better than disconnected models, thus validating the largely presumed importance of modeling recurrent interactions between different cortical regions. In \citet{kiakojouri2020brain}, the authors compare LSSM against trained  multi-layer perceptrons (MLP) in their ability to predict TMS-evoked EEG response and it was found that the introduction of nonlinearity in modeling did not necessarily improve the fit accuracy.
\begin{align}\label{eq:lssm}
    \notag \xbf(t+1) &= \Abf\xbf(t)+\Bbf\ubf(t)+\wbf(t)
    \\
    \ybf(t) &= \Cbf\xbf(t)+\Dbf\ubf(t)+\vbf(t)
\end{align}
Similar to DBS, DCM based on fMRI has also been used to investigate changes in connectivity in response to TMS \citep{grefkes2010modulating,hodkinson2021operculo,pleger2006repetitive}. While in both \citep{grefkes2010modulating} and \citep{pleger2006repetitive} the pre-stimulus and post-stimulus changes in the neural dynamics were analyzed, the work in \citet{hodkinson2021operculo} uses simultaneous TMS and fMRI data to determine changes induced during the application of stimulation. However, all these works are similar in their use of a fixed set of stimulation parameters (frequency and intensity) throughout the entire duration of data acquisition, thus limiting the persistence of excitation and richness of the used data for learning. A summary of TMS-related dynamical system models is presented in Table~\ref{tab:tms}.



\begin{table}[ht]
    \centering
      \resizebox{0.9\textwidth}{!}{%
    \begin{tabular}{|l|l|l|l|l|l|l|}
    \hline
       Model & Data & Paper & $y(t)$ & $u(t)$& \new{Performance} &Order \\
        &  &  &  & & \new{measure} &Selection \\
        \hline\hline
        \multirow{2}{*}{LSSM}&\multirow{2}{*}{Human}&
        \citet{chang2019assessing} & EEG & ON/OFF & MSE &AIC  \\ \cline{3-7}
        &&\citet{kiakojouri2020brain} & EEG & - & RMSE &AIC \\ \hline

        \multirow{3}{*}{DCM}&\multirow{2}{*}{Human}&
        \citet{grefkes2010modulating} & fMRI & - &-&Bayesian  \\ \cline{3-7}
        &&\citet{hodkinson2021operculo} & fMRI & -  & Posterior&Bayesian \\  \cline{3-7}
        &&\citet{pleger2006repetitive} & fMRI & - &- &Bayesian \\ \hline
    \end{tabular}%
    }
    \caption{Summary of data-driven dynamical system models for TMS.}
    \label{tab:tms}
\end{table}




\subsection{Direct Electrical Stimulation}

Direct electric stimulation (DES) is an invasive neurostimulation method similar to DBS which involves direct application of electric current to the brain via surgically placed electrodes. However, unlike DBS where the stimulation is provided in deep subcortical regions (such as the subthalamic nucleus for the treatment of PD), DES is generally limited to either the cortex or the dura mater. Nevertheless, DES has found applications well beyond therapeutic neuromodulation. In particular, DES has been considered a gold standard for functional mapping of the brain for over a decade \citep{mandonnet2010direct, opitz2014validating, mahon2019direct} and has been used widely in identifying regions important for language, somatosensory and motor functions \citep{ojemann1983brain, orena2019investigating,  matsumoto2007functional, ostrowsky2002representation}. Moreover, cortico-cortical evoked potentials (CCEPs) induced in regions distant from the DES site has gained importance as a tool to map the spatio-temporal causal connectivity of the brain \citep{mandonnet2010direct, keller2014mapping}. In brief, CCEP refers to a mapping of the brain's ``effective connectivity" (essentially, transfer function/impulse response) between pairs of cortical regions, where DES is applied to each cortical region and responses (electrocorticography, or ECoG) are recorded from electrodes placed over other cortical regions~\citep{keller2014mapping}. Even though only available for patients with drug-resistant epilepsy undergoing pre-surgical evaluation, CCEPs (and DES in general) provide an invaluable window into the brain's dynamics and have been studied through various approaches, as summarized next.

\subsubsection{Biophysical Models}

A large amount of work describing computational modeling of DES focus on epidural and subdural cortical stimulation. Like DBS, the most common DES response modeling method involves computing the electric field around the stimulation site using FEM \citep{wongsarnpigoon2008computational,seo2016effect}. Commonly, the area around the stimulation including the cortex, grey matter, and dura mater is approximated using a three-dimensional extruded slab model \citep{wongsarnpigoon2008computational, wongsarnpigoon2012computer,vrba2019modeling}. The distribution of the electric potential is then estimated using FEM based on tissue conductivity assumptions and used to determine the de/hyper-polarization of neurons on the cortical surface.  Recent progress in computational techniques has allowed researchers to incorporate more realistic 3D brain models which have been shown to explain the electric field distribution more accurately~\citep{kim2014computational, seo2016effect}.

In \citet{manola2007anodal}, pyramidal neuron CBMs were used in conjunction with a 3D volume conductor model similar to the one in \citep{wongsarnpigoon2008computational} to estimate the electrical field corresponding to the stimulation of the motor cortex. A similar approach to combine neural dynamics with FEM was proposed by \citet{mandonnet2011role}, where a homogenized axonal model was adopted to estimate the regions where action potentials are generated.
However, the stimulation response modeled in all these works is limited to regions near the stimulus site and as such can only be used to understand local de/hyper-polarization activity. While these methods help elucidate functional relevance of specific sites, they have little utility for modeling network-level activities such as CCEPs. Also, FEM calculations often require precise knowledge of electro-geometric properties of cortical regions, even though normative (average) head models can also be assumed.

\subsubsection{Stimulus-Response Models}

As seen in Figure~\ref{fig:dist}, stimulus-response modeling of DES has been pursued as frequently as \rem{the its} biophysical modeling. The former can be roughly divided into two categories based on the repetitiveness of the applied stimulation. First, we have studies that use single-pulse electric stimulation (SPES) to understand the response of the brain under low-frequency DES~\citep{matsumoto2017single,paulk2022local,crocker2021local, valentin2002responses}. Such studies have used temporally well-spaced stimulation pulses (1-5s between consecutive stimulations) and, as a result, evoked activity only in a limited population of localized neurons. In \citep{paulk2022local}, the authors conducted multiple SPES experiments to find the degree of dependency of the evoked responses on stimulation parameters such as input strength (current), distance of the region from the grey-white matter boundary, and the distance of stimulation site from white matter. The results indicated that the responses vary nonlinearly, particularly in regions close to the stimulus site.

In the second category are works that have analyzed the response to a range of low and high-frequency pulses. In studies conducted by \citet{mohan2020effects} and \citet{keller2018induction}, the mapping between DES parameters and responses under repetitive stimulation was modeled using a linear regression models. While these works provide significant insights into the input-output correlation between DES and the expected response, they still have too little temporal resolution for being used in designing closed-loop controllers. Also, attempts to map the brain's response to varying DES parameters without considering the \emph{underlying state} of the brain at the time of stimulation have been shown to lack the capability to model the heterogeneity of responses evoked by DES~\citep{borchers2012direct, papasavvas2020band}.

\subsubsection{Dynamical System Models}

Similar to DBS, several data-driven dynamical system modeling methods have been proposed to examine the brain's spatio-temporal dynamics under DES as showcased in Table~\ref{tab:des}. In \citep{steinhardt2020characterizing}, rectified linear (ReLU) finite impulse response (FIR) models were used to predict the CCEP dynamics at each electrode location. While such an approach does not make use of ``network information" from electrode recordings from other sites, the reasonable level of predictability achieved using this model poses the question of under what circumstances is it important to consider network effects in the system design.

\begin{table}[ht]
    \centering
    \resizebox{\textwidth}{!}{%
     \begin{tabular}{|l|l|l|l|l|l|l|}
    \hline
       Model & Data & Paper & $y(t)$ & $u(t)$& \new{Performance} &Order \\
        &  &  &  & & \new{measure} &Selection \\
       \hline\hline
       MVARX&Human&
        \citet{chang2012multivariate} &iEEG& Stim. current & MSE & CV \\ \hline
        ReLU FIR&Human&\citet{steinhardt2020characterizing}  & iEEG&Stim. current& MSE& - \\ \hline
        \multirow{2}{*}{LSSM}&Simulated&\citet{yang2018control}  &iEEG& [freq., amp.]& RMSE/Bias& AIC \\\cline{2-7}
        &Animal&\citet{yang2021modelling} &LFP power& [freq., amp.]& Cross-correl. & AIC \\ \hline
    \end{tabular}%
    }
    \caption{Summary of data-driven dynamical system models for DES}
    \label{tab:des}
\end{table}

In \citep{chang2012multivariate}, MVARX was instead leveraged to model the DES-evoked activity as measured (invasively) by Stereo-EEG (sEEG) and iEEG. Here, the model input was selected to be the injected current. As in the work of~\citet{steinhardt2020characterizing}, the MVARX modeling here does not restrict the direct flow of input information in the dynamics of non-stimulation channels (as opposed to the more biologically plausible structure where input can \emph{only} flows through the network from the stimulation site). 
However unlike in the FIR model of the former, input information could also enter the model through network interactions.
While the learned model is able to make predictions agreeable with the recorded data, it has to be noted that in this and the majority of similar studies, the input pattern used for the system identification experiment only involved repeatedly stimulating the brain at uniform intervals without varying the current amplitude or waveform shape in any ways, thus limiting the generalizability of the learned models in predicting dynamics under unseen input conditions. However, given that model was learned using real human data and can predict both pre-stimulus and evoked responses makes a promising case for learning more generalizable models. Also, it is worthwhile to mention that the results showcased here indicate that the learned MVARX performs better in comparison to a learned ARX model, thus verifying the often presumed benefits of including network interactions for model predictability.

In \citep{yang2018control}, the authors use binary noise (BN)-modulated stimulation patterns to learn the input-output brain response parameterized using a linear state-space model (LSSM) of the form in~\eqref{eq:lssm} and develop an LQR-based controller for curing neurological conditions, particularly depression, using the developed model. Similar to the DBS-related works of \citet{santaniello2010closed} and \citet{liu2016closed}, the model here was trained using only synthetically generated data. However, instead of using physiologically-derived models (e.g., the basal ganglia network) to generate the simulated input-output data, the authors use a hardware in the loop (HIL)-based simulator to generate ECoG data that simulates real-world delays and noise statistics associated with DBS. Also, the modeling considers the input $\ubf(t)$ to be a vector of the form $[amp(t),freq(t)]$ where $amp(t)$ and $freq(t)$ are the stimulation amplitude and frequency, respectively, and, hence, is generally more explainable from a clinical perspective. 
While the  synthetic nature of the data used here may not accurately correspond to the activity observed in the brain, incorporating realistic delays in the data appears to be a crucial step in developing clinically transferable DES controllers. Additionally, the authors presented a way to transform abstract behavioral objectives (e.g., mitigating depression) to computable control cost functions by relating a mood-based marker with the internal brain state. Although such an approach is yet to be validated using human experiments, it provides a first step towards translations between control-theoretic cost functions and clinical objectives. 

In \citep{yang2021modelling}, the authors extend the same BN-based system identification approach to model the dynamics of large-scale brain networks in non-human primates. The authors compared the predictability of general LSSM against a range of constrained linear models and nonlinear ARX models and it was observed that linear models perform better than the extensively parameterized nonlinear models. Further, among linear models, those which were forced to have non-oscillatory dynamics performed poorly in comparison to unconstrained LSSMs, indicating the importance of oscillatory dynamics in modeling the brain's response to DES. 
Furthermore, to mitigate the often low signal to noise ratio in neural recordings, 
the same BN-stimulation pattern was applied multiple times to average out any noise associated with the internal brain functions.

Considering that data-driven models like MVARX or LSSM require the user to select a wide range of hyper-parameters such as model lags and order, it is important to highlight the tuning methodology followed in the aforementioned models. In \citep{yang2018control} and \citep{yang2021modelling}, the authors use Akaike's information criterion (AIC) to determine the subspace model order. AIC has been widely popular in the modeling literature due to its simple yet effective formulation for balancing model predictability and complexity without the need to use a separate validation dataset. While it has been used in tuning complex models such as ANNs, its application to nonlinear models where one may not clearly define a model order is still a matter of debate. In \citep{chang2012multivariate}, the authors instead define and use a cross-validation (CV) cost function to tune the model lags. While cross-validation methods like k-fold CV are computationally expensive, it is a powerful tool in determining the optimal model order and is a good indicator of model generalizability. In general, various factors such as the model linearity and data stationarity can affect the hyper-parameter selection criteria (AIC, CV, etc.).

\subsection{Transcranial Electric Stimulation}

Transcranial electric stimulation (tES) is a non-invasive neuro-stimulation technology that involves the application of electric currents to the scalp via two or more electrodes. Unlike DBS or TMS, stimulation delivered via tES is weak and is not in general sufficient to evoke action potentials by itself. However, neuronal spiking has been shown to be altered by tES depending on the exact stimulation pattern applied and the timing of the applied stimulation relative to the ongoing brain oscillations~\citep{johnson2020dose,krause2019transcranial,huang2021transcranial,liu2018immediate}. While a large body of tES work focuses on the modulation due to direct current (tDCS), effects concerning the application of alternating current (tACS) and random noise stimulation (tRNS) have also been studied.

\subsubsection{Biophysical Models}

Similar to other neurostimulation techniques, biophysical models have been proposed to understand the effects of tES on the brain. A large number of works approach this problem using a head model including the geometry of the scalp, skull, cortex, and their respective conductivity~\citep{indahlastari2016changing,minhas2012transcranial,sadleir2010transcranial}. Finite element analysis is then used to determine either the current density or the electric field distribution, which can then be used to optimize stimulation parameters. Recent progress in MRI-based anatomical head modeling has allowed for more realistic estimation of the effects in comparison to spherical head models such as the one used in \citet{miranda2006modeling}. This is particularly relevant in tES due to its non-invasive nature where it is critical to model the current flow from the scalp through the skull to the cortex, sub-cortical regions, and the white matter.

Neuronal models like multi-compartment CBMs or neural mass models have been used in the literature to determine the excitatory-inhibitory effects of tDCS \citep{molaee2013effects,arora2021grey, dutta2013neural}. In \citet{arora2021grey}, the authors propose the use of multiple pathways such as modeling tDCS perturbation to ionic gates and ion concentrations to simulate tDCS current density effects. A different approach involving oscillatory neural mass model was used in \citet{kunze2016transcranial} to model the temporal evolution of large-scale brain networks under tES. The parameters of these models (connectivity between regions, for example) are determined through comparison of simulated data against experimental measurements such as EEG and provide interpretable insights into the underlying dynamics. However, these neuronal models still use a highly abstracted representation of the input and do not capture the exact nature of electric conduction from the scalp to individual neurons. More recent works have attempted to alleviate this gap by combining conduction head models with neuronal models, see, e.g.,~\citep{seo2017multi,seo2019relation,rahman2015multilevel}.

\subsubsection{Stimulus-Response Models}

Similar to TMS, stimulus-response analysis of the brain's response to tES has been largely done using non-invasive recordings such as EEG, MEP, and fMRI. In \citet{jamil2017systematic}, the authors monitor pre- and post-stimulation MEP to determine the dependency of cortical excitation on the tDCS intensity. This process of establishing a pre-stimulation baseline and comparison to post-stimulation activity is a recurring mode of analysis in MEP-related literature \citep{wiethoff2014variability,galvez2013transcranial, ammann2017response}. Similar post-hoc studies have been performed using fMRI in works such as \citep{keeser2011prefrontal} where the network connectivity before and after tES are studied. With respect to the simultaneous analysis of response and stimulation, EEG-based monitoring has been leveraged in a large fraction of works \citep{song2014beta,schestatsky2013simultaneous}. Also, a notable pattern in recent works is an increase in the studies of closed-loop tES control design via EEG-based feedback 
\citep{boyle2013eeg,frohlich2021closed,leite2017surface}. For the most part, these works perform on/off control determined by the thresholding of EEG-based bio-markers such as alpha band power \citep{boyle2013eeg} (signal power in the range 8-12Hz). While the control design here is not based on state prediction and optimization, these works shed light on the prospective use of non-invasive signals such as EEG for real-time feedback control of tES.

\subsubsection{Dynamical System Models}

The data-driven dynamical system modeling of brain dynamics under tES is still in its infancy as seen from Table~\ref{tab:tes}. One attempted approach here pertains to the modeling of near-infrared spectroscopy (NIRS)-based hemodynamics taking the EEG recorded during tES (not the tES signal itself) as the input. In \citet{sood2016nirs}, the authors combine ARX modeling with Kalman filters to learn such models online in order for the model to adjust to time-varying data. However, even though EEG can capture the effects of the applied input (stimulation current), such a model cannot directly be used for network control but instead would need to be combined with an inverse mapping from EEG to the actual input current. Nevertheless, the provided proof of concept on the predictability of blood flow activity altered by tDCS using an ARX model is promising.

\begin{table}[ht]
    \centering
        \resizebox{\textwidth}{!}{%
     \begin{tabular}{|l|l|l|l|l|l|l|}
    \hline
       Model & Data & Paper & $y(t)$ & $u(t)$& \new{Performance} &Order Selection \\
        &  &  &  & & \new{measure} &\\
       \hline\hline
       ARX&Human&
        \citet{sood2016nirs} & NIRS& EEG &RMSE  &System ID toolbox (MATLAB) \\ \hline
    \end{tabular}%
    }
    \caption{Summary of data-driven dynamical system models for tES}
    \label{tab:tes}
\end{table}
\subsection{Optogenetics}
Broadly, optogenetics refers to the combination of genetics and optics to control well-defined events within any specific cell type of living tissue, including neurons~\citep{yizhar2011optogenetics}. Commonly, this is achieved via a viral insertion of genes into (neuronal) cells that endow them with light responsiveness and the (invasive) delivery of light to the affected neurons. Notably, this process can be finely controlled to generate both excitatory and inhibitory effects in the targeted neurons (even multiple interconnected populations of neurons within the same brain region) depending on the inserted genes, timing of gene insertion, and the used wavelength of light.

Even though not yet approved for use in humans, optogenetics plays a unique role in the applications of control theory in neuroscience.
Indeed, one of the main goals of developing network control models of the form in~\eqref{eq:ss} is control design, where it is often desirable to tap into the wealth of classical control methods and readily apply them to neural dynamics. However, doing so is challenging due to the great mismatch between the level of \emph{spatiotemporal specificity} required by most control methods and that provided by existing neurotechnologies. This makes optogenetics a particularly attractive stimulation technology for control design, where cell-type specific spatial resolution and millisecond temporal resolution are combined~\citep{deisseroth2006next}. 

\subsubsection{Biophysical Models}

A common approach to the modeling of neural responses to light stimuli involves detailed modeling of the physics underlying optogenetics. Briefly, optogenetics involves the opening of light-sensitive ion channels or pumps which leads to ionic currents similar to the natural currents on the right hand side of~\eqref{eq:cbm}. In other words, each channel/pump has multiple, say $\ell$ internal states $s_1, \dots, s_\ell$ (at least two: open and close) whose dynamics change under the influence of light. Therefore, a popular family of models for optogenetic effect~\citep{grossman2011modeling,foutz2012theoretical,stefanescu2013computational,nikolic2013computational,selvaraj2014open} start from a CBM of the form in~\eqref{eq:cbm} and augment it into
\begin{subequations}\label{eq:opto-s}
    \begin{align}
        C_i \frac{d}{d t} v_i(t) &= I_i(t) + \sum_{j = 1}^n \bar g_{ij} p_{ij}(t) [E_j - v_i(t)] + g_{\rm opsin}(t) [E_{\rm opsin} - v_i(t)],
        \\
        g_{\rm opsin}(t) &= \left(\sum_{k = 1}^\ell s_k(t) g_k \right) \phi(v_i(t)),
        \\
        \frac{d}{dt} \sbf(t) &= \Kbf(u(t), t) \sbf(t).
    \end{align}
\end{subequations}
Here, $g_{\rm opsin}(t)$ is the light-dependent conductance of this channel/pump, $E_{\rm opsin}$ is the constant equilibrium potential of the optogenetically activated channel/pump (depending on the gene type and light wavelength used), $s_k(t)$ is the fraction of channels/pumps on the surface of the neuron in state $k$ (so $s_1 + \cdots + s_\ell = 1$), $g_k$ is the conductance of the channels/pumps if they were all in state $k$, $\phi(\cdot)$ is an optional nonlinear function that can model the voltage-dependence of $g_{\rm opsin}$, $u(t)$ is the light intensity, and $\Kbf$ is a matrix of state transition rates which depends on various linear or nonlinear forms on $u(t)$ and, possibly, time. Note that even though the light intensity signal is often a sequence of square pulses in implementation, $u(t)$ is often taken as an impulse train
\begin{align}
    u(t) = \sum_k \delta(t - t_k)
\end{align}
where the times $t_k$ correspond to the onset times of the square pulses in light intensity.
If the internal states $s_1, \dots, s_\ell$ 
are not important, \eqref{eq:opto-s} can be simplified by simply using a linear dynamical system to directly model the mapping from $g_{\rm opsin}(t)$ to $u(t)$~\citep{witt2013controlling,shewcraft2020excitatory,vierling2010computational}, often using a 1st-3rd order system with an impulse response similar to~\eqref{eq:ltiir}.

Such biophysical models have been used to build closed-loop feedback controllers of optogenetic input. One strategy that has been demonstrated is to let $u(t)$ \emph{modulate} certain parameters of the optogenetic input, such as the amplitude of a light intensity square-wave pulse while fixing other parameters such as frequency and duty cycle~\citep{srinivasan2018closed}. A more sophisticated version was used in~\citep{newman2015optogenetic}, where a single control input $u(t)$ in the model determines the parameters of two simultaneous sources of optogenetic stimulation, modulating the frequency, pulse width, and intensity envelope of one source (ChR2, excitatory) while directly mapping to the light intensity (as in a naive model) of the second source (eNpHR3.0, inhibitory).

Another class of methods involves using spiking network models. In \citet{valverde2020deep}, exponential integrate-and-fire neurons were used to build a simplified motor cortex network with around 1000 neurons. While having access to all neuronal activities at a microscopic level provides great modeling flexibility and fidelity, it can be difficult to interpret such microscale models without resorting to aggregate statistics. As such, \citet{mahrach2020mechanisms} instead use population-level modeling, where connectivity between 4 neuronal populations was tuned to best explain experimental data.

\subsubsection{Stimulus-Response Models}
Analysis of response to optogenetic input has largely been done using fMRI-based monitoring of rodent brain activity~\citep{liang2015mapping,kahn2013optogenetic,cover2021whole}. For the most part, works using such opto-fMRI studies can be summarized as attempts to either profile BOLD signals in relation to the application of stimulation or map changes in functional connectivity. In~\citet{krawchuk2020optogenetic}, the authors go a step further and model the dependencies between cerebral blood flow and photo-stimulation parameters (pulse duration, frequency) using linear regression. Optogenetic response has also been studied in the context of non-human primates in works such as~\citet{yazdan2018targeted} and \citet{bloch2022cortical}. While both the works study the changes in network connectivity induced by stimulation, the approach in~\citet{bloch2022cortical} is particularly interesting as the authors explicitly model the connectivity between regions as a non-linear (polynomial) function of selected features including the anatomical location of the regions, their distance from the stimulus and the distance between considered regions.

\subsubsection{Dynamical System Models}

System identification-based approaches for response modeling of optogenetic stimulation have been proposed as highlighted in Table~\ref{tab:opt}. In \citet{bolus2020state}, the authors demonstrated the ability of LSSM towards the prediction of neuronal firing rates in response to the intensity of input light. A distinguishing feature in this work is the addition of a ``disturbance" term in the state model to account for drifts in the neural response, which was shown to significantly impact the estimation accuracy. Besides the prediction performance, the work here is particularly interesting as the model learned here was shown to be effective in modulating the firing activity of a single neuron in a live rat through an LQR-based controller.

\begin{table}[ht]
    \centering
        \resizebox{\textwidth}{!}{%
     \begin{tabular}{|l|l|l|l|l|l|l|}
    \hline
       Model & Data & Paper & $y(t)$ & $u(t)$& \new{Performance} &Order \\
        &  &  &  & & \new{measure} &Selection \\
       \hline\hline
       LSSM&Animal&
        \citet{bolus2020state} &Firing rate& Stim. intensity & $R^2$ & - \\ \hline
        \multirow{4}{*}{DCM~/~MDS}&\multirow{4}{*}{Animal}&\citet{pinotsis2017linking}  &LFP&- & Posterior&Bayesian \\\cline{3-7}
        &&\citet{bernal2017studying} &fMRI& Stim. ON/OFF  & Posterior&Bayesian \\ \cline{3-7}
        &&\citet{grimm2021optogenetic} &fMRI& Stim. ON/OFF  & Posterior&Bayesian \\ \cline{3-7}
        &&\citet{ryali2016combining} &fMRI& Stim. ON/OFF  & Posterior&Bayesian \\ \hline
    \end{tabular}%
    }
    \caption{Summary of data-driven dynamical system models for Optogenetics}
    \label{tab:opt}
\end{table}

Consistent with the modeling literatures of DBS and TMS, dynamic causal modeling has also been employed in optogenetic modeling. In \citet{pinotsis2017linking}, a DCM based on real LFP data was used to study the effective connectivity changes between input on and off conditions. In a related work~\citep{ryali2016combining}, the authors propose the use of multivariate dynamical systems (MDS) modeling, a data-driven biophysical approach similar to DCM, to train separate models corresponding to stimulation on and off conditions. While this is one way to model the effects of the input, response can also be modeled by explicitly considering the stimulus as a driving input of the underlying brain network as seen in the DCM-related works of \citet{bernal2017studying} and \citet{grimm2021optogenetic}. In both works, DCM based on fMRI data was used to quantify the causal relationship between brain regions such as the subthalamic nucleus and motor cortex. While the response to optogenetic input is what is being modeled in these works, the analysis of effective connectivity in these regions can also potentially facilitate the understanding of the mechanism of other neurostimulation techniques such as subthalamic DBS in this case. Ultimately, the versatility of learned models can perhaps be regarded as an important factor in building a generalizable model of the brain.

\section{Disease specific modeling approaches}

\new{In the above sections, we categorized the modeling literature based on the neurostimulation technique and the approach employed. While we discuss the use of these models briefly in areas such as control and functional mapping, the real appeal of these works lies in the treatment of neurological and psychiatric conditions. To that end, here we provide a complementary categorization of the literature based on the brain-related disorders considered and highlight important trends in building models geared towards treating specific diseases using stimulation.}

\new{Parkinson's Disease (PD) marks one of the most widely studied disorders in the context of neurostimulation. It is a neurological disorder related to abnormalities in basal ganglia (BG) and associated reductions in dopamine production~\citep{little2015computational}. Naturally, a range of biophysical models of BG have been proposed to study the PD mechanism under free-response (i.e., no external input) and stimulation conditions~\citep{popovych2012desynchronizing,little2015computational}. Several PD-focused review articles have synthesized this literature, see, e.g., \citet{chia2020historical} and \citet{shimohama2003disease}. A majority of the neurostimulation models of PD employ biologically-derived BG networks to evaluate the connectivity changes related to PD and stimulation. A wide range of modeling detail and network size has nevertheless been used, from model with a relatively small number of neurons ($\approx$10)~\citet{terman2002activity} to denser network structures comprising of more neurons ($\approx$500) aimed at accurately modeling the neuronal firing and oscillatory activity~\citet{hahn2010modeling}. For the most part, these works focus on DBS as the mode of stimulation which comes as no surprise considering the proven effectiveness of DBS in treating PD.}

\new{A more control-theoretic description of the neuron population dynamics perhaps comes from the use of Kuramoto models \citep{kuramoto1975self} in works such as \citep{lysyansky2011desynchronizing,holt2016phasic}. While several variations of the Kuramoto model has been used, these works all seek to model the phase synchronization in the firing activity of neuron populations observed in patients with PD. Despite relying on assumptions such as homogeneity of neuron sub-systems and the type of coupling between then, the obtained models have proven useful in their ability to abstract the complex neural dynamics of PD into simple phase equations and hence allowing the user to define relatively simple objectives for control design. Furthermore, several dynamical system models discussed earlier were originally constructed with the precise intent of building closed loop controllers for the treatment of PD \citep{liu2016closed,su2018nonlinear,haddock2017model,kahan2014resting,kahan2019deep}. While a number of these works have used simulated datasets which use some of the aforementioned biophysical models as the ground truth in order to learn and validate linear models \citep{liu2016closed,su2018nonlinear}, works such as \citep{haddock2017model,kahan2014resting} have used real human data to study the same.}

\new{Epilepsy refers to a group of chronic neurological diseases characterized by unprovoked
seizures~\citep{wendling2016computational,wang2015computational} and marks another historical focus of neurostimulation. Unlike PD where the symptoms are rather persistent, the onset of epileptic seizures can be difficult to predict and hence modeling the dynamics under epilepsy requires investigating the mechanism that causes seizures and conditions that lead to their termination~\citep{wendling2016computational}. Similar to PD, biophysical neural mass models have been employed in the literature to mimic the activity seen during the seizure duration commonly known as the ictal phase~\citep{goodfellow2012self, mina2013modulation, fisher2013deep}. In \citet{wilson2014hamilton}, the authors use a single-neuron CBM to build a minimum-time controller to guide the brain to a target set of desirable state in the smallest possible time. In \citet{mina2013modulation}, the authors instead use a multi-compartment CBM of the cortex and the thalamus to model the LFP response during seizure and stimulation conditions using real human data. Besides biophysical modeling, data-driven large scale modeling of the brain seen in the works of \citet{chang2012multivariate}, \citet{chang2020model}  and \citet{steinhardt2020characterizing} is another approach in the pursuit of modeling the epileptic brain under stimulation.}

\new{In general biophysical models often need to precisely model the regions and biological interactions underlying the specific disorder of interest~\citep{wang2015computational} which can in turn enhance their appeal for the treatment of that disorder while limiting their applicability to other conditions. Data-driven modeling, on the other hand, is often more universal in methodology and disorder-specific modeling often becomes a matter of using the correct input-output data $u(t), y(t)$ and choosing the appropriate parametric model structure~\citep{yang2018control}.
This has allowed for the recent use of dynamical system models for modeling the effect of neurostimulation in a diverse class of disorders \citep{liu2016closed,chang2012multivariate,yang2018control,gibson2017impact}. Besides the neurological disorders mentioned in the above paragraphs, this data-driven approach has also been extended to psychological conditions, such as the use of LSSM to model the activity of a DES-stimulated brain under depression in \citet{yang2018control} and the use of DCM-based connectivity models in the context of obsessive compulsive disorder (OCD) in \citet{gibson2017impact}.}

\section{Discussions}

In this paper we reviewed a large body of work on the computational modeling of the brain's response to five of the most commonly used neurostimulation technologies. For each technology, we categorized the surveyed works into three categories of biophysical, stimulus-response, and (data-driven) dynamical system modeling. Perhaps not surprisingly, similar biophysical models have been employed across neurostimulation techniques which, for the most part, use finite-element volume conduction models, conductance-based neuron models, or a combination thereof. \new{In general, biophysical models are computationally intensive and may scale poorly to large-scale brain networks. This is due to their reliance on iterative finite element calculations which are well-known to be computationally expensive~\citep{giudice2019analytical} and/or detailed multi-neuron dynamics which would inevitably face a curse of dimensionality if combined to form large-scale brain networks}. However, this complexity comes with the advantage of having a dynamical model that describes the neural activity at the fine and biologically transparent level of synaptic exchanges and neuron depolarization and hyper-polarization, leading to the popularity of these models in the literature~\new{\citep{popovych2012desynchronizing, salvador2011determining,lysyansky2011desynchronizing,shirinpour2021multi}}. 

The second category of reviewed works for each stimulation technology consisted of statistical stimulus-response models. 
While these models may not provide insights into the low-level neuronal response to neurostimulation, they instead allow for the abstraction of the complex network activity in terms of summary biomarkers of interest (connectivity, band power, tremor, etc.). The main limitation of this category of works lies in their need for large sample sizes and 
repeated experiments which may be costly to obtain, particularly in clinical settings~\new{\citep{munoz2020researcher,akbar2020mapping,yarossi2019experimental}}. 

Lastly, we have dynamical system models as the main focus of the present review which take the data-driven aspect of the stimulus-response models one step further. By using time series data, these models learn temporal dependencies in input-output data via dynamical models. Here, a large body of works has used linear models such as ARX, MVARX, and LSSM and have indeed been shown to achieve good prediction accuracy despite the nonlinearity of the brain at the microscale~\new{\citep{yang2021modelling,nozari2020brain}}. In addition to lower computational complexity in model fitting and validation, the use of linear models has far-reaching implications for control design and the interpretability of network dynamics given the wealth of knowledge on the analysis and control of linear dynamical systems~\new{\citep{kailath1980linear,anderson2007optimal}}. 
We further reviewed several works that have approached nonlinear dynamical system modeling of neurostimulation and have shown the power of ANNs, nonlinear ARX, and DCMs based on one-step-ahead prediction objectives. Ultimately, data-driven dynamical system models, whether linear or nonlinear, are critical for control theoretic methods to find their potentially widespread application in neural engineering and therapeutics.

Data-driven modeling comes with its own pros and cons. The theory-driven nature of biophysical models and the many assumptions that underlie them often leaves a considerable amount of variance in real data unexplained~\new{\citep{nozari2020brain}}. 
On the other hand, data-driven models of an appropriate order and complexity can reduce the unexplained variance through model fitting. 
However, learning data-driven models, particularly at scale, requires the availability of extensive and distributionally rich input-output data. The latter, in particular, necessitates carefully designed data collection experiments with persistently exciting stimulation waveforms which are currently uncommon and costly. 
The problem of acquiring rich datasets is further exacerbated in clinical settings by the fact that events such as seizures are unpredictable and rare. This in turn contributes to the sparsity of samples corresponding to simultaneous occurrences of stimulation and medically relevant conditions and events.

The great potential of system identification and dynamical system models in neuroscience is paralleled with major limitations in the existing literature and challenges to be tackled in future works. These limitations and challenges stem in part from the many differences that exist between the brain and the engineering systems for which system identification and control theory have been classically developed~\new{\citep{ziegler1942optimum,kirk1970optimal,qin2003survey}}. These include dimensionality and the number of constituent subsystems, stochasticity and maximum achievable predictive accuracy, the number of spatio-temporal scales relevant for system behavior, and technological limitations in sensing and actuation, to name a few. As such, different modeling and control techniques need to be developed for the brain. Moreover, and in part due to the mentioned challenges, the vast majority of the dynamical system models developed for the brain have used either simulated data or real data in a post-hoc manner, i.e., without demonstrating the utility of the learned models for control design. As such, we expect a combination of custom-designed modeling and control algorithms and iterative, system identification-informed experiments to be necessary (and perhaps sufficient) for unlocking the great potential of data-driven control design for neurostimulation.

\section*{Conflict of Interest Statement}

The authors declare that the research was conducted in the absence of any commercial or financial relationships that could be construed as a potential conflict of interest.

\section*{Author Contributions}

Conceptualization: SR, EN. Literature synthesis: GA, EN. Writing: all.

%
%
%

\bibliographystyle{Frontiers-Harvard}
\bibliography{local}





\end{document}